\documentclass[11pt,a4paper,reqno]{amsart}

\usepackage{siunitx}
\usepackage{graphicx} 
\usepackage[headings]{fullpage}
\usepackage{amsaddr}
\usepackage{dsfont}
\usepackage[singlelinecheck=false]{caption}
\usepackage{url}
\usepackage{multirow}
\usepackage{mathtools}
\usepackage{upgreek}

\newcolumntype{Y}{>{\centering\arraybackslash}X}

\newcommand{\R}{\mathbb{R}}
\newcommand{\Z}{\mathbb{Z}}
\newcommand{\cx}{c_{\textup{1}}}
\newcommand{\cy}{c_{\textup{2}}}
\newcommand{\cz}{c_{\textup{3}}}

\newcommand{\gramm}{\mathrm{g}}

\newcommand{\Wraw}{W^{\text{raw}}}
\newcommand{\Gfg}{G^{\text{fg}}}
\newcommand{\Gscaled}{G^{\text{scaled}}}
\newcommand{\Graw}{G^{\text{raw}}}
\newcommand{\Gmorph}{G^{\text{morph}}}
\newcommand{\Geroded}{G^{\text{eroded}}}

\begin{document}
\title[]{An AFM-based approach for quantification of guest particle deformation during mechano-fusion}

\author[]{Phillip Gräfensteiner$^1$, Judith Friebel$^2$, Lisa Ditscherlein$^2$, Orkun~Furat$^1$, Urs A. Peuker$^2$, Volker Schmidt$^1$}
\address{$^1$Institute of Stochastics, Ulm University, 89069~Ulm, Germany}
\address{$^2$Institute of Mechanical Process Engineering and Mineral Processing,\\ Technische Universität Bergakademie Freiberg, 09599 Freiberg, Germany}

\begin{abstract}
During the mechano-fusion process for dry particle coating, (hetero)-aggregates are formed consisting  of host particles which are coated by smaller guest particles. During this process, the latter are exposed to intense particle--particle interactions and particle--wall impacts, which lead to deformation of the guest particles original shape. These deformations on the nano- and microscale can heavily influence the effective macroscopic properties of the resutling coated particles. We present a method to quantify the shape deformation of guest particles during mechano-fusion based on measurements acquired by atomic force microscopy before and after mechano-fusion. To this end, we reconstruct the 3D shape of guest particles by means of an ellipsoidal fit, constrained by the known volume of the guest particles. Using these reconstructed shapes, we can quantify the degree of deformation by comparing the aspect ratios of the  ellipsoidal fits before and after mechano-fusion. Such a quantification enhances the understanding of how process-related parameters influence the geometric descriptors of the involved particles, which in turn impact the overall macroscopic properties of the material.
\end{abstract}

\maketitle

\section{Introduction}
\begin{figure}[ht]
    \centering
    \includegraphics[width=7cm]{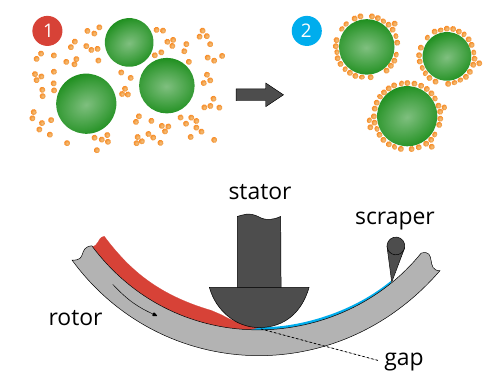}  
    \caption{Simplified scheme of the mechano-fusion process. Top: Host (green) and guest  (orange) particles before (1) and after mechano-fusion (2). Bottom: Setup of high-intensity mixer consisting of rotor, stator and scraper.}
    \label{fig:MechanoFusion}
\end{figure}

Mechano-fusion (MF) is a process for dry particle coating. Having been developed and studied in the 1980s~\cite{Yokoyama.1987}, the process is experiencing a reemergence of importance through applications in pharmaceuticals~\cite{Kumon.2008, Stank.2014} and functionalized battery materials manufacturing~\cite{Zheng.2019b, Phattharasupakun.2021}.
Host particles, with particle sizes in the range of micrometers, are coated with guest particles~\cite{Pfeffer.2001}. The size of the guest particles can range from the low nanometer up to the low micrometer scale. Coating takes place in a high-intensity mixer with a rotor-stator setup~\cite{Jay.2006}. A process scheme is shown in Figure~\ref{fig:MechanoFusion}.

The coating process is initiated by  (i) the input of mechanical energy during mechano-fusion by means of centrifugation of particles to the rotor wall, (ii) compaction of the particles’ powder bed inside the small gap between rotor and stator, and (iii) the collision of particles after the scraper zone~\cite{Lau.2017}.
As a result, particle deformation can occur for some material systems due to intense particle-particle interactions and particle-wall impacts. Excess mechanical energy dissipates in the form of thermal energy. Despite the use of active cooling, the process chamber and the materials can heat up during the process~\cite{Chavda.2013}, which can contribute to particle deformation.
In previous studies, polymeric materials, like polymethylmethacrylate (PMMA), polystyrene (PS) or polytetrafluoroethylene (PTFE), have been used as host or guest particles during MF~\cite{Chou.2008}. 
Polymers, like other materials, are characterized by an initially elastic behavior that turns into plastic deformation under increasing stress. Additionally, time- and temperature-dependent viscoelastic behavior has been observed in polymers~\cite{Brostow.2007}.
A recent study with a model material system consisting of non-polymeric alumina host particles and polymeric PS guest particles showed that the PS particles are deformed during MF under certain process parameters. 
This deformation leads to changes of the coating's microstructure, which in turn not only influences microscopic particle--particle interactions, but also the macroscopic properties of the coated particles as a bulk material, like flowability~\cite{Luddecke.2021}.
To investigate how process-related parameters impact macroscopic properties, it is crucial to gain a quantitative understanding of the influence of these parameters on geometric descriptors, such as the degree of deformation.

In the present paper, we consider an approach for quantitatively assessing the degree of deformation in guest particles after MF in comparison to pristine guest particles before MF based on
surface topography data acquired by atomic force microscopy (AFM). 
AFM is part of the family of scanning probe microscopy technologies and is widely used for micro- and nanoscale surface analysis of specimens in materials science because of the high achievable resolution, which allows imaging of features in the nanometer range. 
In AFM, the probe is a cantilever with known defined spring constant. AFM can be used under two different modes of operation, namely contact-mode (C-AFM) and non-contact mode (NC-AFM). In C-AFM the cantilever is in direct contact with the sample surface. During line-wise scanning the cantilever's height, \emph{i.e.}, its $z$-position, is feedback controlled via a set point force for acquiring topography images~\cite{Haugstad.2012}. In our case, the use of C-AFM is suitable in the quantification of the deformation behavior. Further processing of such topographic images has to be performed in order to quantify the shape and thus the structural deformation of guest particles.

Our approach for quantifying the deformation of guest particles is based on fitting an ellipsoid to the topographical data of the measured surface of each guest particle under a side constraint, which is obtained by exploiting knowledge of the volume of the guest particles. 
With the resulting reconstructions of the 3D shapes of the guest particles at hand, we can analyze the shape deformation by considering various geometrical descriptors, see~\cite{furat.2021} for an overview. In particular, we focus on the aspect ratio, \emph{i.e.}, the ratio of major-axis to minor-axis lengths, of the reconstructed ellipsoidal shapes before and after MF.
In this way it is possible to quantify the deformation that occurs during the mechano-fusion process by means of a single scalar descriptor.
Such a quantification opens the door for a better understanding of process-microstructure relationships, as it allows to study how process-related parameters influence the geometry of the involved particles. 
To the best of our knowledge, so far, only limited studies have been performed on 3D surface modeling of data acquired by AFM measurements~\cite{shilo.2004}.

\section{Materials and Methods}\label{sec:MaterialAndImaging}
\subsection{Material system}\label{sec:Materials}
\begin{figure}[ht]
    \centering
    \includegraphics[width=\textwidth]{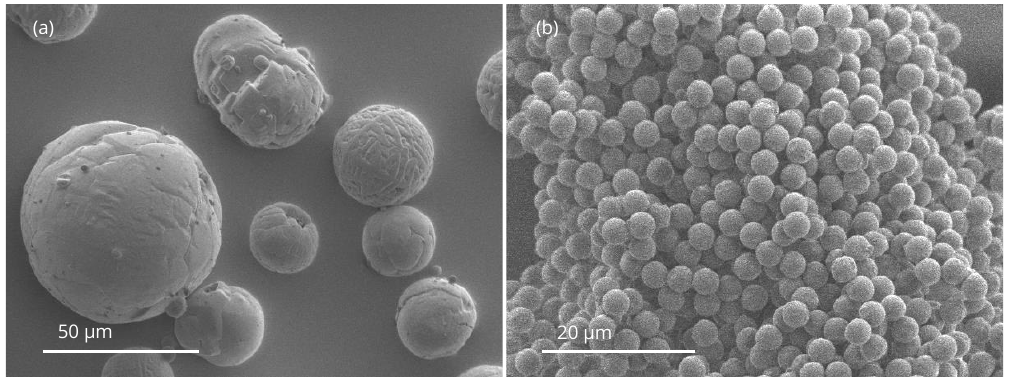}  
    \caption{SEM images visualizing the considered particles. Alumina acts as host particle substance (a) and PS as guest particle substance (b).}
    \label{fig:SEM-before}
\end{figure}

Acting as host particle substance, spherical alumina (Al$_{2}$O$_{3}$) with a density of \SI{3.95}{\gramm / \centi\metre\cubed} is obtained from Denka Chemicals (Düsseldorf, Germany). The median particle size of the host particles is $d_{50}=\SI{45.64}{\micro\metre}$. In addition, strongly crosslinked spherical and monodispersing PS is used as guest particle substance. PS under the trade name Chemisnow\textsuperscript{\textregistered} SX-350H is provided by Kowa Europe (Düsseldorf, Germany) and has a density of \SI{1.05}{\gramm / \centi\metre\cubed} and a particle diameter of \SI{3.5}{\micro\metre} as specified by the manufacturer. Scanning electron microscopy (SEM) images of the pristine host and guest particles as received by the above mentioned manufacturers are shown in Figure~\ref{fig:SEM-before}.

\subsection{MF process}\label{sec:MF-process}
The coated particles investigated in this paper stem from the dry particle coating process of mechano-fusion implemented into the picobond module of the Hosokawa Alpine (Augsburg, Germany) small scale multi-processing platform picoline. For a targeted monolayer coating, the mass ratio of guest particles in the feed of the MF process can be calculated for monodispersing and ideally spherical host and guest particles using a formula given in~\cite{Yang.2005}. An estimation for the given particle system results in a mass ratio of 9.12\,wt\% of PS. From initially three performed experiments, the one with the most particle deformation as suggested by SEM images is chosen to be investigated here. For the experiment of interest, the MF process is carried out under ambient conditions, at a rotational speed of 5000\,rpm for 10\,minutes. A gap size of \SI{2}{\milli\metre} is used between rotor and stator. During the process, the temperature increase on the outer surface of the process chamber lid is monitored and reaches \SI{7.01}{\kelvin}. 

\begin{figure}[ht]
    \centering
    \includegraphics[width=\textwidth]{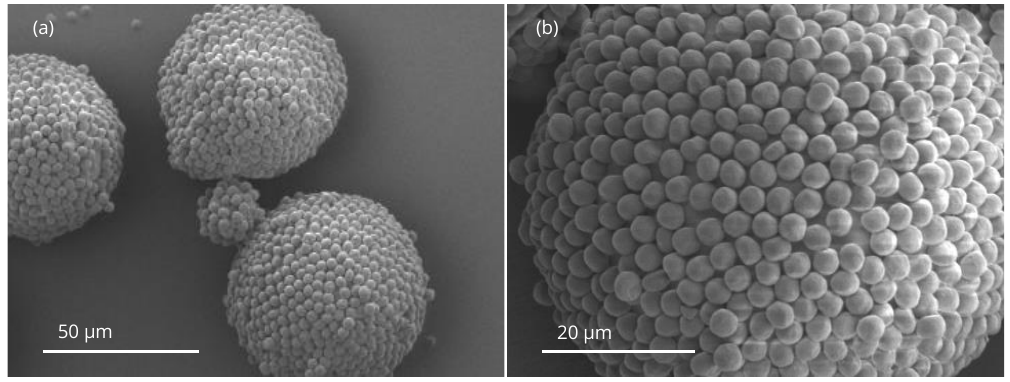}  
    \caption{SEM images of coated particles acquired at the two different magnifications of 300x~(a) and 1000x~(b).}
    \label{fig:SEM-after}
\end{figure}

During mechano-fusion, the effect of an increase in temperature on the deformation of guest particles can be superimposed to purely mechanically induced plastic deformation~\cite{Naito.1993}.
For polymers and other amorphous materials, the so-called glass transition temperature is the temperature where structural changes take place on the molecular level and the material begins to flow viscously~\cite{Dyre.2006}. This is a relevant property to consider when discussing deformation and viscoelastic behavior, where the latter refers to all non-elastic, time- and temperature-dependant deformation behavior, see Chapter 13 in~\cite{vanKrevelen.2009}. 
For PS, the glass transition occurs at around \SI{100}{\degreeCelsius}~\cite{Rieger.1996}. Despite the temperature increase inside the process chamber, it is assumed here that the PS guest particles are processed below the glass transition temperature. We assume that in this temperature range, the
elastic and ideal plastic deformations of PS guest particles are characterized by volume preservation. This assumption is, for example, used in models for describing heavily deformed spherical contacts~\cite{Wadwalkar.2010} and, in our case, plays a crucial role in modeling the guest particle deformation.
After the coating process, the material is removed from the process chamber. Small sub samples are created by using a micro-rotary sample splitter. Figure~\ref{fig:SEM-after} shows SEM images acquired of a sample of coated particles after splitting. The alumina particles are mostly coated with a monolayer of PS particles. By visual inspection, deformation of the guest particles is observed when comparing guest particles in Figure~\ref{fig:SEM-after} with the pristine guest particles in Figure~\ref{fig:SEM-before}b.

\subsection{AFM scanning}
\begin{figure}[ht]
    \centering
    \includegraphics[width=11cm]{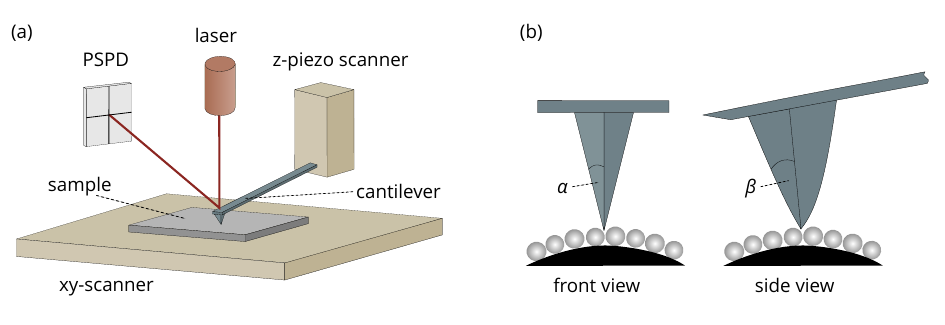}  
    \caption{(a) Simplified scheme of the AFM measuring setup for C-AFM. (b) Front and side view of cantilever tip geometry on the surface of the coated particle. The alumina particle is shown in black, and the PS particles  in gray. The side wall angles of the cantilever tip are indicated by $\alpha$ and $\beta$.}
    \label{fig:AFMscheme}
\end{figure}

The AFM XE-100 by Park Systems Europe GmbH (Mannheim, Germany) is used in this study. During C-AFM the cantilever is brought into close contact with the sample surface under constant force. A laser is focused on the tip of the cantilever. Subsequently, the deflection of the laser is registered on a position-sensitive photodetector (PSPD). More precisely, the deflection of the cantilever, caused by the interaction of the cantilever with the sample surface, influences the deflection angle of the laser. The PSPD can consequently provide information on the sample's $z$-position, \emph{i.e.}, the height at the measured position. During line-wise scanning of the surface, the $z$-position of the cantilever is additionally feedback controlled by a $z$-piezo scanner to allow the cantilever to traverse features with greater $z$-expansion of the specimen surface. A 3D surface profile is the result of the C-AFM topography scan. 
It is important to note that the $z$-values measured by AFM are only meaningful in relation to each other, as AFM can only measure changes in height relative to the point where the cantilever first contacts the sample. At this first point of contact, a $z$-value of $0$ is recorded. All other measured $z$-values then represent the change in height relative to that point.
A scheme of the AFM scan setup is presented in Figure~\ref{fig:AFMscheme}.

\begin{figure}[ht]
    \centering
    \begin{tabular}{ccccl}
    \frame{\includegraphics[width=0.19\textwidth]{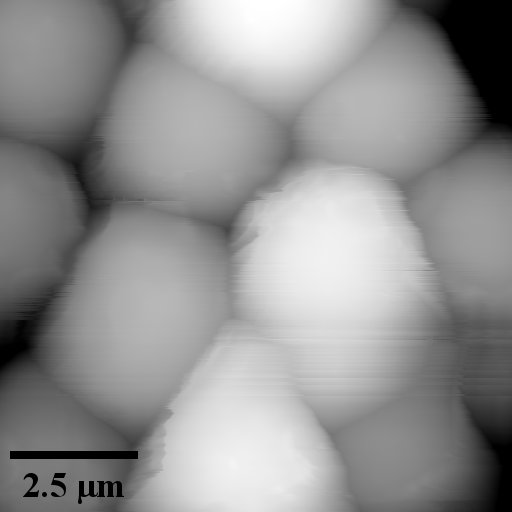}} &
    \frame{\includegraphics[width=0.19\textwidth]{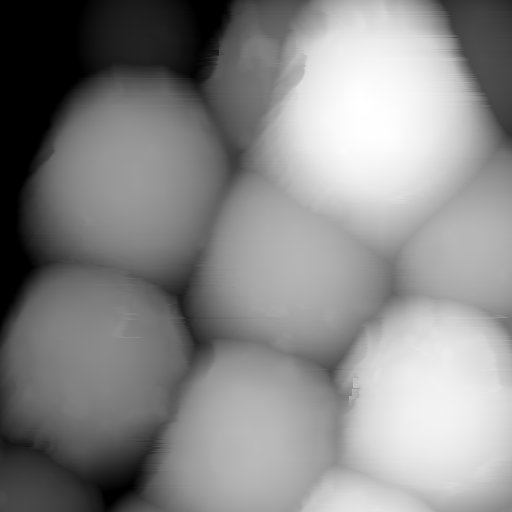}} &  
    \frame{\includegraphics[width=0.19\textwidth]{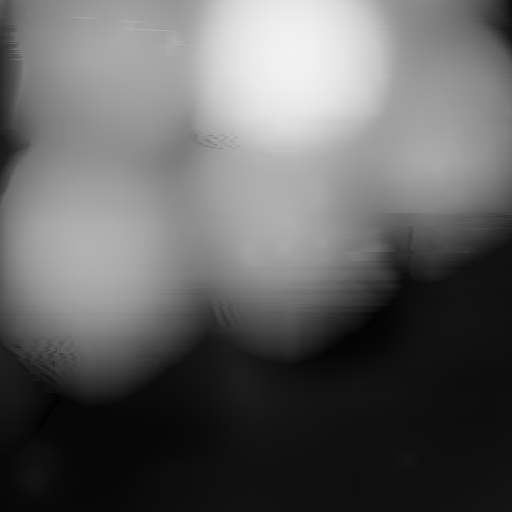}} &
    \frame{\includegraphics[width=0.19\textwidth]{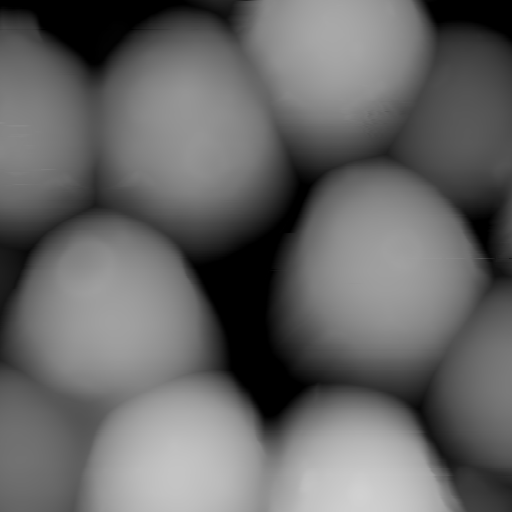}} &\multirow{2}{*}{\includegraphics[trim=3ex 0 0 4.5cm, width=0.1\textwidth]{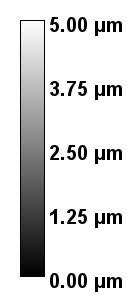}}\\
    \frame{\includegraphics[width=0.19\textwidth]{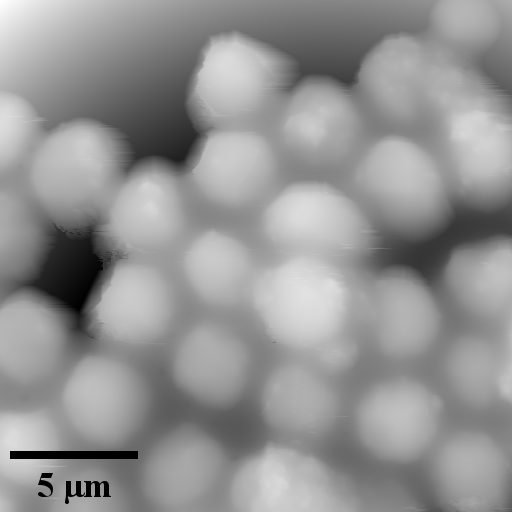}} &  
    \frame{\includegraphics[width=0.19\textwidth]{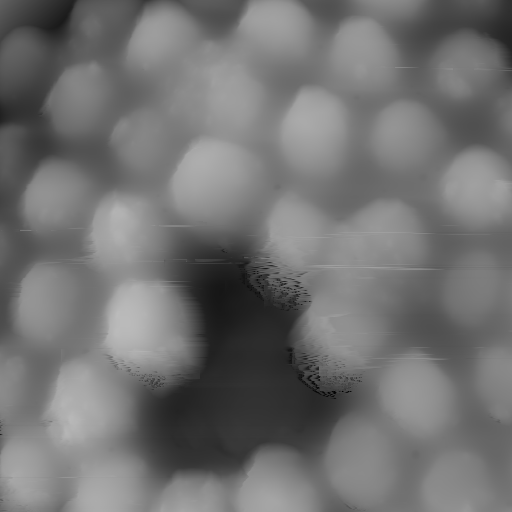}} &
    \frame{\includegraphics[width=0.19\textwidth]{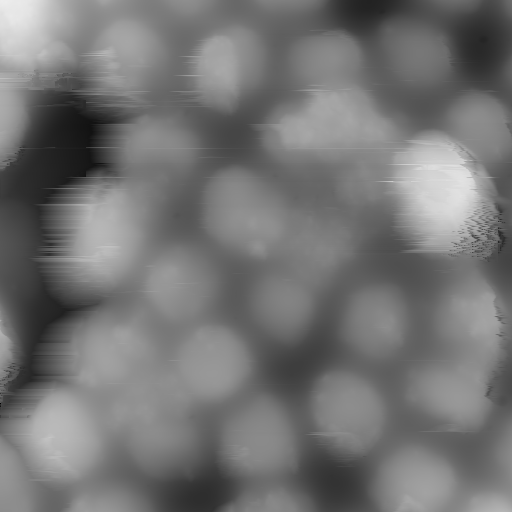}} &
    \frame{\includegraphics[width=0.19\textwidth]{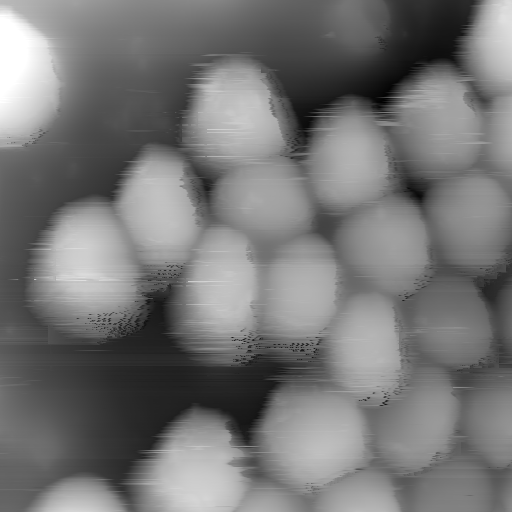}} &
    \end{tabular}
    \caption{Gray-scale visualizations of AFM measurements. Top row: Pristine PS  particles before MF. Bottom row: PS particles after MF. The gray-scale values of each image indicate the measured height relative to the respective lowest
observed point of that measurement.
}
    \label{fig:greyscales}
\end{figure}

As part of the sample preparation, silicon wafers are cleaned using acetone followed by Milli-Q water. Subsequently, a two-part epoxy resin is thinly distributed on top of silicon wafers. The sample of coated particles is dispersed on the epoxy film of the first wafer. Similarly, a sample of pristine PS particles is dispersed on the epoxy film of the second wafer. The samples are placed to harden over night.
Topography scans in this study are performed using contact mode with a standard, monolithic silicon cantilever of type ContAl-G by Budget Sensors (Sofia, Bulgaria). The tip of the cantilever is very sharp ($<\SI{10}{\nano\metre}$). For samples with larger variability in $z$-direction and sharp edges on the surface, the resolution of the scan is reduced due to the sidewall angle of the conically shaped tip. 
Artifacts produced by the cantilever tip geometry and the sample surface are referred to as tip-sample convolution effects~\cite{Shen.2017}. Here, the space between two guest particles is defined by a high $z$-expansion and steepness. In these areas, the resolution can be limited and the correct $z$-value of the  point in the space between guest particles can be inaccessible to the cantilever tip.
The side walls of the used cantilever are inclined \SI{20}{\degree} to \SI{25}{\degree} along the cantilever axis (angle $\alpha$ in Figure~\ref{fig:AFMscheme}b) and \SI{25}{\degree} to \SI{30}{\degree} perpendicular to the cantilever axis (angle $\beta$ in Figure~\ref{fig:AFMscheme}b). At the apex the inclination is \SI{10}{\degree}.
The force set point is chosen to be small (\SI{9}{\nano\newton}) to prevent scratching of PS particles with the sharp tip. The scan area is either 10 or \SI{20}{\micro\metre}. Exemplary gray-scale images of the particle topographies are shown in Figure~\ref{fig:greyscales}.

\subsection{Curvature of host particles}\label{sec:flattening}
Flattening is a common procedure during AFM image processing. Images are prone to show slope or curvature effects as a result of the mechanical setup of the sample and scanner stages (e.g. tilted sample, drift during scanning)~\cite{Wang2018}. Before quantitative analysis of features, a removal of those artifacts by flattening is necessary to retrieve the underlying background. This prevents features from being analyzed in distorted shape. In our case, the curvature of the host particles (median particle size is \SI{45.64}{\micro\metre}, see Section~\ref{sec:Materials}) is an additional factor. In order to correctly extract information about the shape of the guest particles from the AFM measurements of coated host particles, we need to adjust the data to compensate for the curvature of the spherical host particle.

\begin{figure}[h]
    \centering
    \begin{tabular}{cc}
    \includegraphics[width=0.45\textwidth]{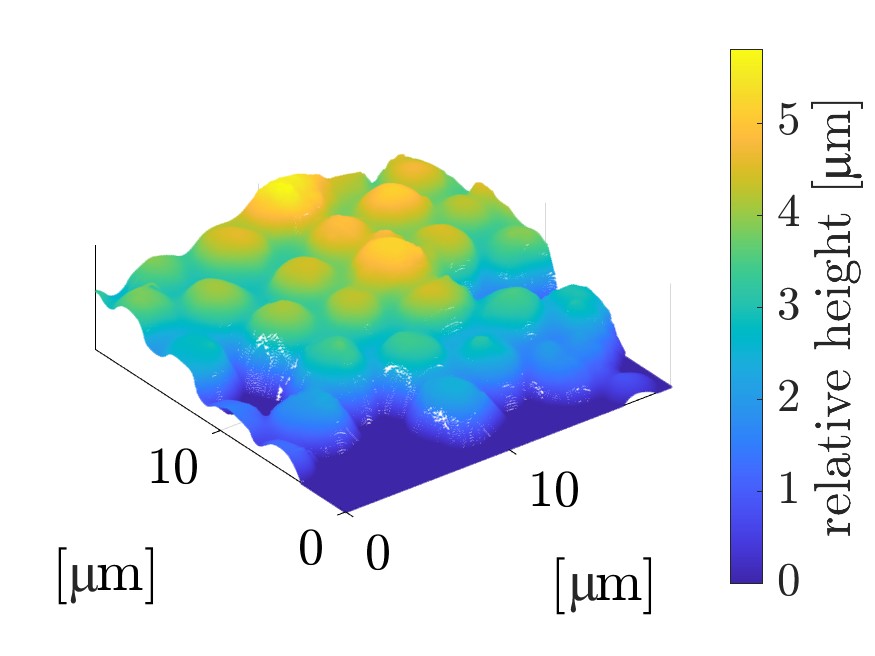} &
        \includegraphics[width=0.45\textwidth]{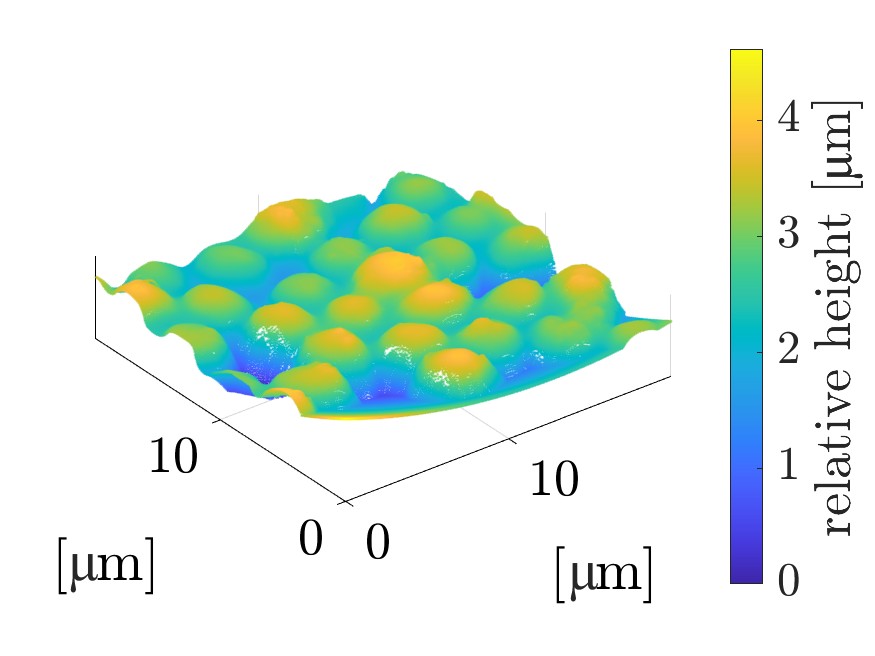}\\
    (a) & (b)
    \end{tabular}
    \caption{Visualization of an AFM measurement before~(a) and after~(b) flattening. The color indicates the measured height relative to the respective lowest observed point of that measurement.}
    \label{fig:curvature}
\end{figure}

The flattening is performed on the raw images of coated alumina particles using the software XEI (version 4.1.1) from Park Systems, where 
a polynomial function of second order is fitted to the overall curvature along the $x$-axis and subsequently along the $y$-axis.
More precisely, we first perform an averaging of the measured $z$-values along the $y$-axis, after which a polynomial of second order is fitted to the resulting averaged $1$-dimensional profile. The corresponding values of this polynomial are then subtracted from all data points in order to remove the curvature in $x$-direction. 
This procedure is then repeated analogously in order to also remove the curvature in $y$-direction.
See Figure~\ref{fig:curvature} for an example of a measurement before and after removing the curvature of the host particle.

\subsection{Particle-discrete segmentation}\label{sec:segmentation}
In the next step we extract regions within AFM measurements which correspond to individual guest particles, so that they can be analyzed separately.
The approach involves applying a marker-based watershed algorithm to the image data acquired by AFM measurements.
The image data are treated as $16$-bit gray-scale images in this subsection, where each pixel corresponds to a data point of the measured surface. As the data points were collected over a square grid, the location of each pixel directly corresponds to the $x$- and $y$-value of the data point. 
The gray-scale value of each pixel corresponds to the $z$-value measured at the location of that pixel, as seen in Figure~\ref{fig:greyscales}. 
Formally, we consider a gray-scale image $G$ to be a mapping from the infinite domain $\Z^2=\{\ldots,-1,0,1,\ldots\}\times \{\ldots,-1,0,1,\ldots\}$ to $[0,1]$. Though real images are only defined over a finite subset $D_G\subset\Z^2$, they can be extended in an appropriate sense, usually by assigning a constant value to pixels outside of $D_G$, to fit into this framework. Analogously, binary images are considered as mappings from $\Z^2$ to $\{0,1\}$.

\begin{figure}[ht]
    \centering
    \begin{tabular}{ccc}
        \frame{\includegraphics[width=0.25\textwidth]{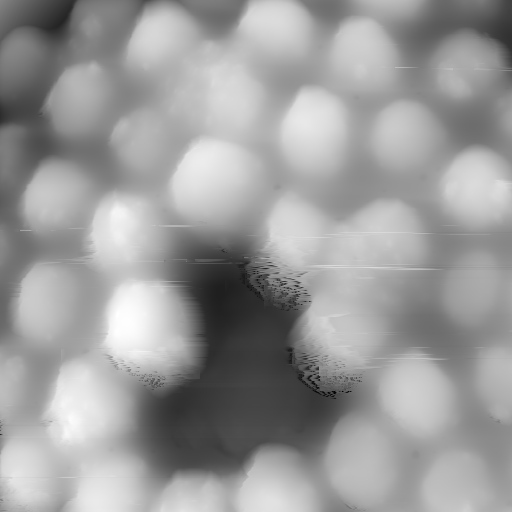}} &
        \frame{\includegraphics[width=0.25\textwidth]{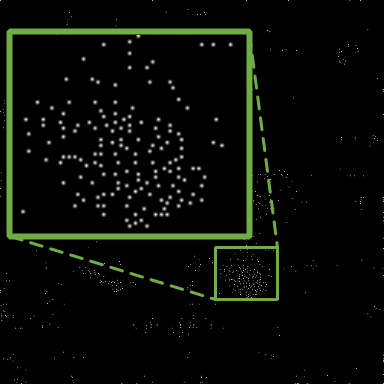}} &
        \frame{\includegraphics[width=0.25\textwidth]{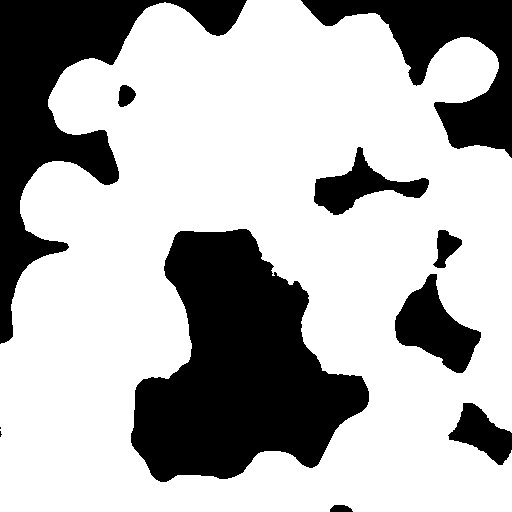}}\\
        (a) & (b) & (c) \\
        \frame{\includegraphics[width=0.25\textwidth]{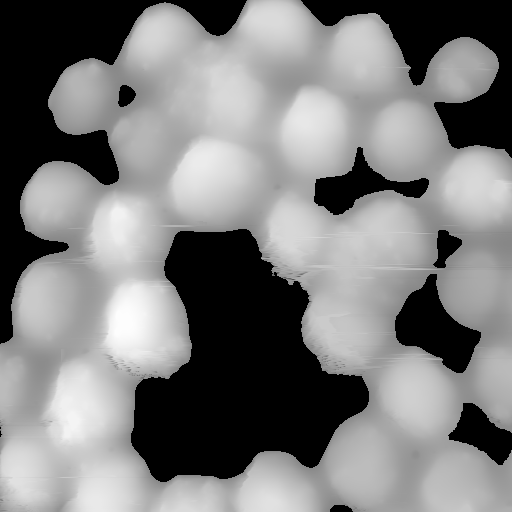}} &
        \frame{\includegraphics[width=0.25\textwidth]{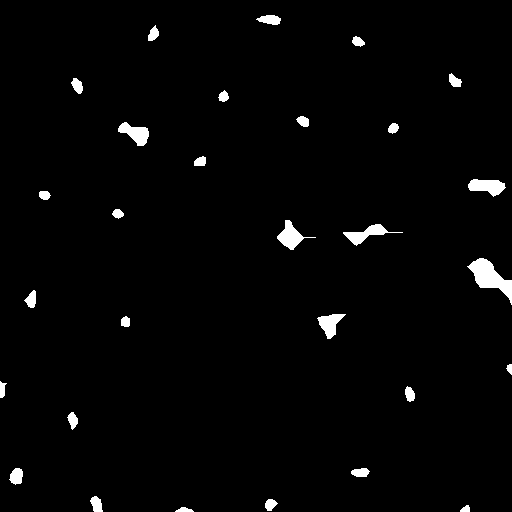}} &
        \frame{\includegraphics[width=0.25\textwidth]{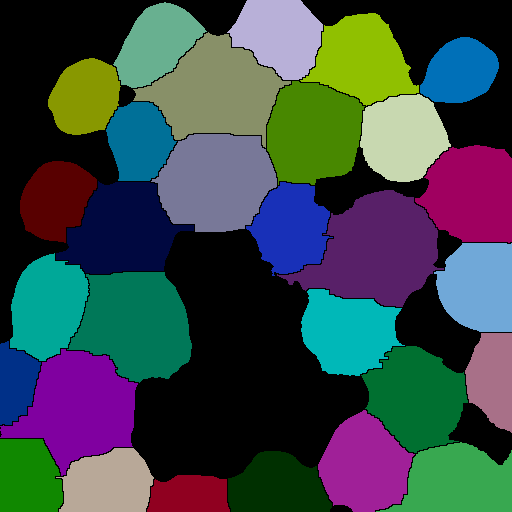}}\\
        (d) & (e) & (f)
    \end{tabular}
    \caption{(a) Gray-scale image $\Graw$ of AFM measurement. The gray-scale values of each pixel contain information about the height measured at the corresponding location. (b) Binary image showing regional maxima of $\Graw$. A small region is magnified for visualization. (c) Binary mask $F$ used to label the foreground of $\Graw$. (d) Gray-scale image $\Gmorph$ after preprocessing. (e) Marker image $M$ used for the watershed algorithm. Connected components indicate the locations of guest particles. (f) False color image of the particle-discrete segmentation $S$ resulting from the marker-based watershed algorithm.}
    \label{fig:segmentation}
\end{figure}

We demonstrate the necessary preprocessing steps at the example of a raw $16$-bit gray-scale image $\Graw$ acquired by an AFM measurement, see Figure~\ref{fig:segmentation}a. At first, the values are shifted and scaled so that they are within the range of the unit interval $[0,1]$. That is, we consider the scaled image $ \Gscaled$ given by
\begin{equation}\label{eq:scaling}
    \Gscaled(x,y)=\big(\Graw(x,y)-\min_{(x,y)\in D_G}\Graw(x,y)\big)/\big(\max_{(x,y)\in D_G}\Graw(x,y)-\min_{(x,y)\in D_G}\Graw(x,y)\big),
\end{equation}
for eeach $(x,y)\in\Z^2$.
The classical watershed algorithm is based on the idea of interpreting the gray-scale image $G$ as a topographical map.
Intuitively speaking, water sources are placed at regional minima (alternatively, regional maxima after inverting the image by multiplying pixel values with $-1$) and the topographical map is flooded with water rising from these sources.
The lines, along which the resulting water basins stemming from different water sources meet, induce a segmentation of the original image. Crucially, the number of distinct regions within the obtained segmentation is equal to the number of initial water sources.
As the data exhibits a large amount of regional maxima, see Figure~\ref{fig:segmentation}b, a direct application of the watershed algorithm to the gray-scale images would lead to drastic oversegmentation.
This issue can be remedied by adaptations of the classical watershed algorithm. 
In particular, in a marker-based watershed algorithm the user can specify the location and shape of initial water sources by means of a binary marker image, where each individual connected component within the binary image is considered to be its own water source. We determine this binary marker image so that each connected component corresponds to the location of a guest particle in the gray-scale image data. For more information on classical and marker-based watershed algorithms, see~\cite{meyer.1990,roerding.2003}.

Since some images contain large areas in which no guest particles are present, we create a binary mask $F$ that labels the foreground of each image using manual thresholding, \emph{i.e.}, a generous outline of the area where guest particles are visible, see Figure~\ref{fig:segmentation}c.
More precisely, we determine $F$ from the topography image $\Gscaled$ by
\begin{equation}
    F(x,y) = \begin{cases}
    1,\quad \text{if } \Gscaled(x,y)\geq t,\\
    0,\quad \text{else},
    \end{cases}
\end{equation}
for each $(x,y)\in\Z^2$, where $t\in[0,1]$ is the manually chosen threshold.
This mask is then multiplied pixel-wise with the gray-scale images, so that background pixels have the value $0$, which leads to the foreground image $\Gfg$ given by
\begin{equation}
    \Gfg(x,y)=\Gscaled(x,y)F(x,y),
\end{equation}
for each $(x,y)\in\Z^2$.
In order to determine the binary marker image, we perform  morphological opening by reconstruction on the gray-scale image, which reduces the amount of regional maxima, see Section~6.3.9 in~\cite{soille.2003}. This process consists of two steps, namely first performing a morphological erosion using some structuring element $E\subset\mathbb{Z}^2$, followed by morphological reconstruction of the eroded image, using the original image as a mask. We explain these steps in more detail. For morphological erosion, the gray-scale value of the eroded image $\Geroded=\Gfg\ominus E$ at  location $(x,y)\in\Z^2$ is given by the minimum value within the overlap of $E$ shifted by $(x,y)$ and $\Gfg$, that is,
\begin{equation}\label{eq:erosion}
    \Geroded (x,y) = \min_{(s,t)\in E}\Gfg(x+s,y+t),
\end{equation}
for each $(x,y)\in\Z^2$.
Here we choose the structuring element $E$ to be a disk centered at $(0,0)$ with a radius of $5$--$15$ pixel, depending on the resolution of the measurement under consideration. Recall that for real image data that is only defined on a finite subset of $\mathbb{Z}^2$, we assume that the image is extended in an appropriate sense onto the whole grid $\mathbb{Z}^2$. 
When computing the erosion, an extension by the value $\max_{(x,y)\in D_G}G(x,y)$ is desired in order to avoid edge effects when taking the minimum in Eq.~\eqref{eq:erosion}. In our case, this corresponds to the value $1$, as the images are already scaled by Eq.~\eqref{eq:scaling}. If a maximum is considered, then an extension by the value $0$ is more appropriate.
The here considered notion of morphological erosion can be extended to structuring elements that are themselves gray-scale images, see Section~3 in~\cite{soille.2003}.

The second step of morphological opening by reconstruction  is the morphological reconstruction, which is performed on the eroded image $\Geroded$, using the uneroded gray-scale image $\Gfg$ as a mask. 
This involves iterative morphological dilation $\oplus$, defined analogously to Eq.~\eqref{eq:erosion} with $\max$ instead of $\min$, where we choose the structuring element $E$ to be a centered square of $3$ pixel side length. 
After each iteration, the pixel-wise minimum between the dilated image and the mask image $\Gfg$ is taken. This process stops once the image remains unchanged during an iteration.
Formally we set $\Gmorph_0 = \Geroded$ and repeat
\begin{equation}
    \Gmorph_k = \min(\Gfg, \Gmorph_{k-1}\oplus E)
\end{equation}
until $\Gmorph_k=\Gmorph_{k-1}$ for some integer $k\geq1$.
The resulting image $\Gmorph$ contains a greatly reduced amount of regional maxima, whose locations are primarily at the center of guest particles, see Figure~\ref{fig:segmentation}d.
Examples of further applications of morphological reconstruction on gray-scale images can be found in~\cite{vincent.1992}.

We consider the binary image $B$, whose foreground consists of the regional maxima of $\Gmorph$, \emph{i.e.}, the connected components of pixels with a constant value, that are surrounded by pixels with a lower value.
As some guest particles still contained more than one connected component of regional maxima, we perform a morphological closing on the binary image $B$.
This involves a morphological dilation, followed by a morphological erosion, where the structuring element is chosen as a disk centered at $(0,0)$ with a radius of $20$--$30$ pixel, depending on the resolution of the given measurement. 
This causes multiple connected components, corresponding to the same guest particle, to merge into a single connected component. The resulting image $M$ is the binary marker image used for the marker-based watershed algorithm, see Figure~\ref{fig:segmentation}e.

The gray-scale image before morphological opening by reconstruction $\Gfg$ is now modified so that it only exhibits regional maxima which coincide with the connected components in the binary marker image $M$. This step also involves morphological reconstruction, see Section~6.3.6 in~\cite{soille.2003} for details. Now, an application of the watershed algorithm (to the inverted image) floods the topography starting from water sources that correspond to the locations of guest particles. The result of this algorithm is a particle-discrete segmentation $S\colon\Z^2\rightarrow \{1,2,\ldots\}$, which is a labelled image with integer values, where pixels with the same value correspond to the same region.
Note that the watershed algorithm partitions the set of pixels. That is, regions resulting from flooding of a source provided by the marker image might contain pixels associated to the background. Therefore, the value of all background pixels is set to 0 using the binary mask $F$ that labels the foreground, see Figure~\ref{fig:segmentation}f.
The procedure was implemented in Matlab~\cite{MATLAB}, which contains functions to perform various tasks involving morphological operations, including morphological reconstruction.

For the remainder of this manuscript, we only consider regions that do not contact the boundary of the image data.
This is to eliminate edge effects where a particle was only partially measured, which would not lead to a meaningful analysis of the corresponding measurement.

\subsection{Reconstruction of guest particle shapes by ellipsoidal fits} \label{sec:ellipsoid_fit}
The particle-discrete segmentation procedure described above divides the image data into regions, each containing information about the measured exposed surface of a single guest particle.
In order to reconstruct their full shape in 3D, we assume that the guest particles are ellipsoidal.
This assumption is motivated by the fact that the pristine guest particles before coating, \emph{i.e.}, before mechanical stress is applied to them, are nearly spherical, so that deformation by unidirectionally applied forces would naturally lead to an ellipsoidal shape.
Since information about the curvature of the host particle has already been removed from the image data as described in Section~\ref{sec:flattening}, we also assume that one principle axis of the measured guest particles is aligned with the $z$-axis. The surface of 
such an ellipsoid is then given by those $\mathbf{x}\in\R^3$ fulfilling the equation
\begin{equation}
    \label{eq:ellipsoid_def}
    (\mathbf{x}-\mathbf{c})^\top\mathbf{R}_\gamma^\top\mathbf{AR_\gamma}(\mathbf{x}-\mathbf{c})=1,
\end{equation} 
where $\mathbf{c} = (\cx,\cy,\cz)\in\R^3$ are the coordinates of its center, $\mathbf{A}=\textup{diag}(a,b,c)$ is a diagonal $3\times3$ matrix whose diagonal entries $a,b,c>0$ are the lengths of its three semi-axes, and 
\begin{equation}\label{eq:rot_mat}
    \mathbf{R_\gamma}=\begin{pmatrix}
\cos \gamma & -\sin \gamma  & 0\\
\sin \gamma & \cos \gamma   & 0 \\
0           &  0            & 1
\end{pmatrix}
\end{equation} is a matrix that describes the rotation by the angle $\gamma\in[0,2\pi)$ around the $z$-axis. Note that the semi-axis that is aligned with the  $z$-axis has length $c$. Moreover, the rotation matrix $\mathbf{R}_\gamma$ is applied to the shifted vector $\mathbf{x}-\mathbf{c}$, so that the center $\mathbf{c}$ of the ellipsoid belongs to the axis of the rotation described by $\mathbf{R}_\gamma$.
In order to achieve a reliable and realistic fit, we also make use of information about the size of guest particles. More precisely, we assume that their volume does not change during the coating process, as explained in Section~\ref{sec:MF-process}.
Since the pristine guest particles are nearly spherical and monodisperse with a radius of \SI{1.75}{\micro\metre}, the ellipsoidal fit should have a volume of $\frac{4}{3}\pi (\SI{1.75}{\micro\metre})^3$. 

Let $G$ be the gray-scale image corresponding to a measurement acquired by AFM, and $S$ the corresponding particle-discrete segmentation  obtained as described in Section~\ref{sec:segmentation}. Furthermore, let $A$ be a region of $S$, \emph{i.e.}, a non-empty set  of the form 
\begin{equation}
   A = \{(x,y)\in\Z^2\colon S(x,y)=k\},
\end{equation} 
for some $k\in\{1,2,\ldots\}$. In the following, it will be more convenient to consider the data provided by an AFM measurement as a set of data points in $\R^3$ rather than as a gray-scale image.
Hence, consider the discrete set of points $V\subset\R^3$ obtained by the AFM measurement corresponding to the image $G$, with values given in micrometer. Let $\Wraw\subset V$ denote the subset of those points whose $x$- and $y$-values correspond to pixels within the chosen region $A$ of $S$. Then, with $\nu>0$ denoting the resolution of the AFM measurement in $x$- and $y$-direction, we have
\begin{equation}
    \Wraw = \big\{(\nu x,\nu y,z)\in\R^3\colon (x,y)\in A, z=G(x,y)\big\}.
\end{equation}
Here we only require a scaling of the $x$- and $y$-component, since, as opposed to Section~\ref{sec:segmentation}, no shifting and scaling was performed on the gray-scale image $G$, so that the $z$-values are already given in micrometer.
It is convenient to first perform a general centering of the set $\Wraw$, by shifting its center of gravity in the $(x,y)$-plane to its origin $(0,0)$. For this purpose, we put
\begin{align}
    g_1 &=\frac{1}{\vert \Wraw\vert} \sum_{(x,y,z)\in \Wraw}x\\
    g_2 &=\frac{1}{\vert \Wraw\vert} \sum_{(x,y,z)\in \Wraw}y,
\end{align}
where $\vert \Wraw\vert$ denotes the cardinality of the set $\Wraw$. 
We then shift $\Wraw$ by $(g_1,g_2)$ in the $(x,y)$-plane, which leads to
\begin{equation}
    W=\big\{(x-g_1,y-g_2,z)\colon (x,y,z)\in \Wraw\big\}.
\end{equation}
Now, consider the loss function $H\colon\R^3\times(0,\infty)^3\times[0,2\pi)\longrightarrow [0,\infty)$ given by 
\begin{equation}
    \label{eq:loss_function_def}
    H(\cx,\cy,\cz,a,b,c,\gamma)=\sum_{w\in W}\Vert w - P(w)\Vert^2,\qquad \cx,\cy,\cz\in\R,a,b,c>0,\gamma\in[0,2\pi),
\end{equation}
where $\Vert\,\cdot\,\Vert$ denotes the Euclidean norm on $\R^3$ and $P(w)$ is the orthogonal projection of $w\in\R^3$ onto the surface of the ellipsoid described by Eq.~\eqref{eq:ellipsoid_def}.
In order to compute $P(w)$, we use the algorithm implemented in~\cite{fileexchange.ellipsoid}.
In other words, the function $H$ is the sum of squared orthogonal distances between the measured data points of a single guest particle and the surface of the ellipsoid given by the parameter vector $(\cx,\cy,\cz,a,b,c,\gamma)$.
Thus, we can fit an ellipsoid to the data points in $W$ by minimizing $H$. Since we have prior knowledge on the guest particle volume of $\frac{4}{3}\pi R^3$ with $R=1.75$, we minimize $H$ using a constraint for volume conservation of the fitted ellipsoids, \emph{i.e.}, by
\begin{equation}
    \label{eq:min_problem_general}
    \begin{aligned}
    &\min_{\cx,\cy,\cz,a,b,c,\gamma} &H(\cx,\cy,\cz,a,b,c,\gamma)\\
    &\text{subject to}    &\frac{4}{3}\pi( abc-R^3) = 0.
    \end{aligned}
\end{equation}

\begin{figure}[ht]
    \centering
    \begin{tabular}{ccc}
    \includegraphics[width=.3\textwidth]{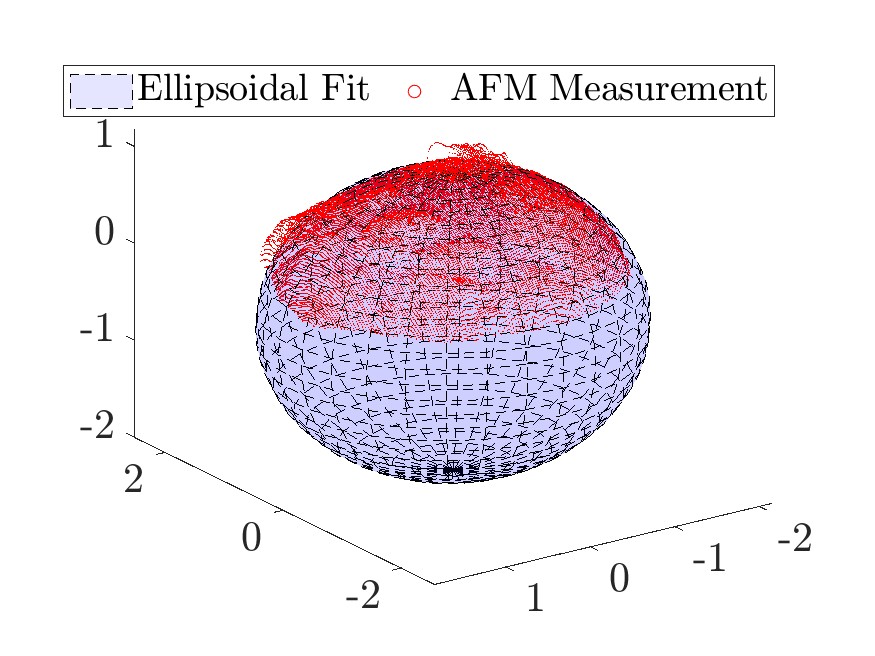} & 
    \includegraphics[width=.3\textwidth]{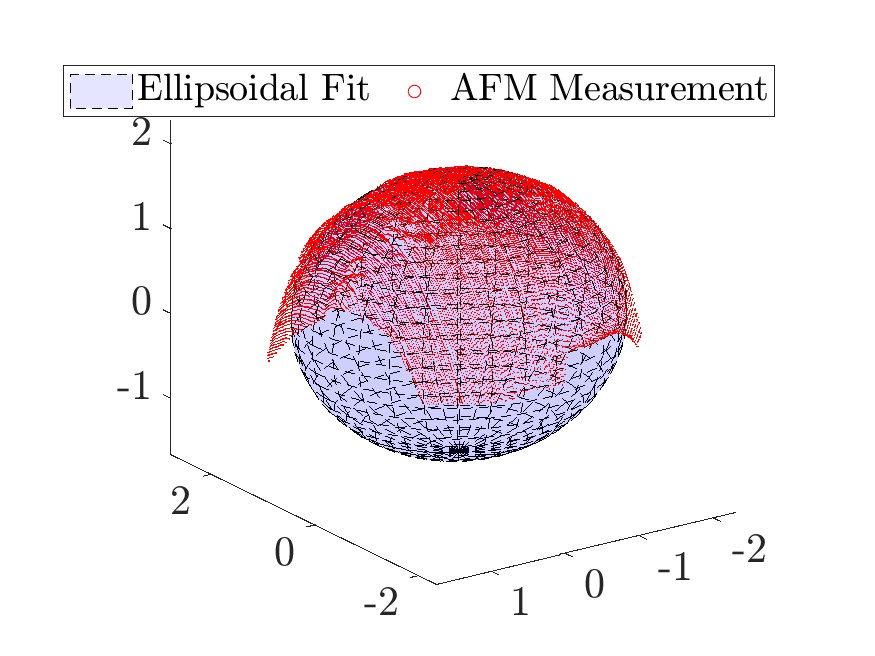} &
    \includegraphics[width=.3\textwidth]{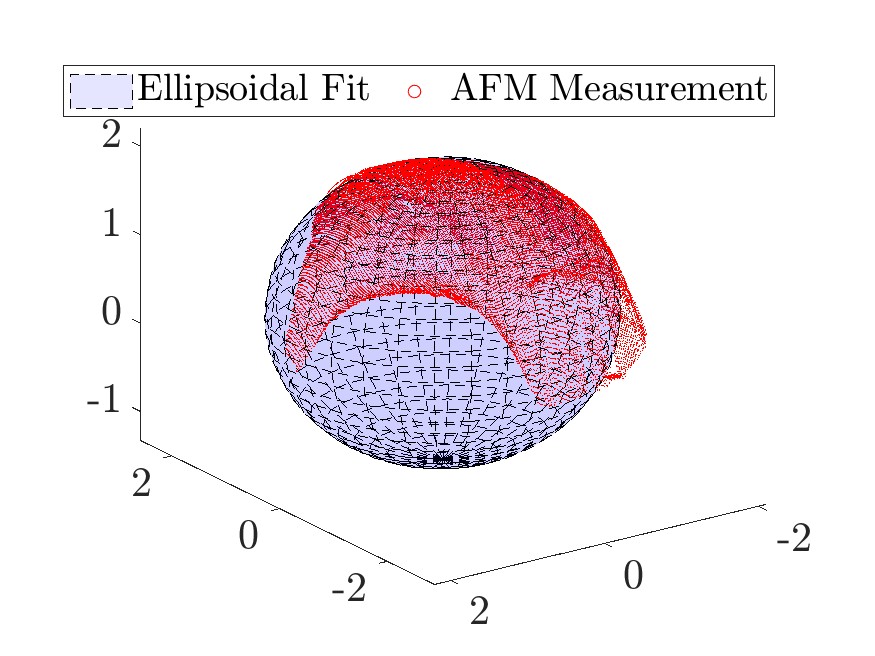} \\
    \includegraphics[width=.3\textwidth]{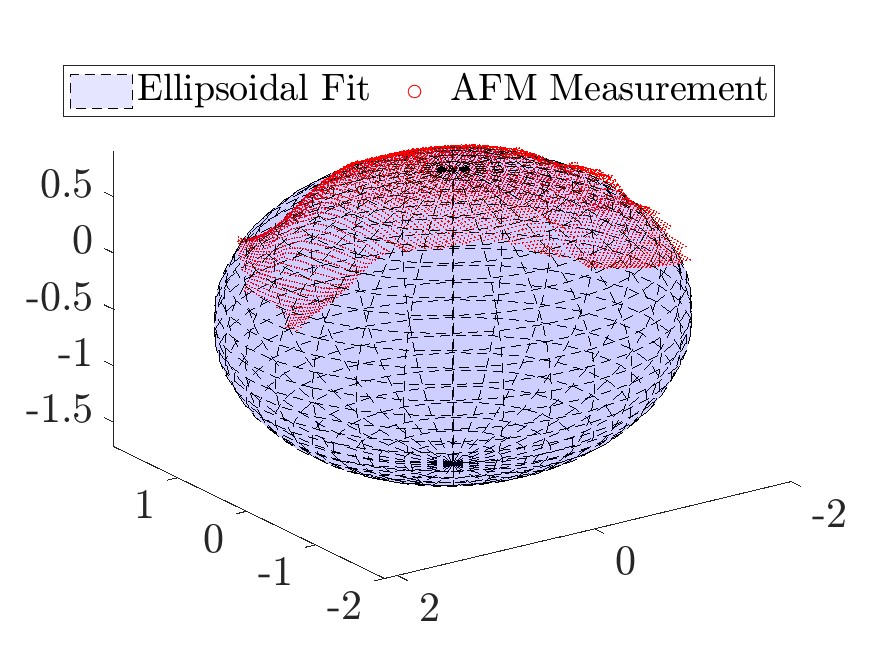} & 
    \includegraphics[width=.3\textwidth]{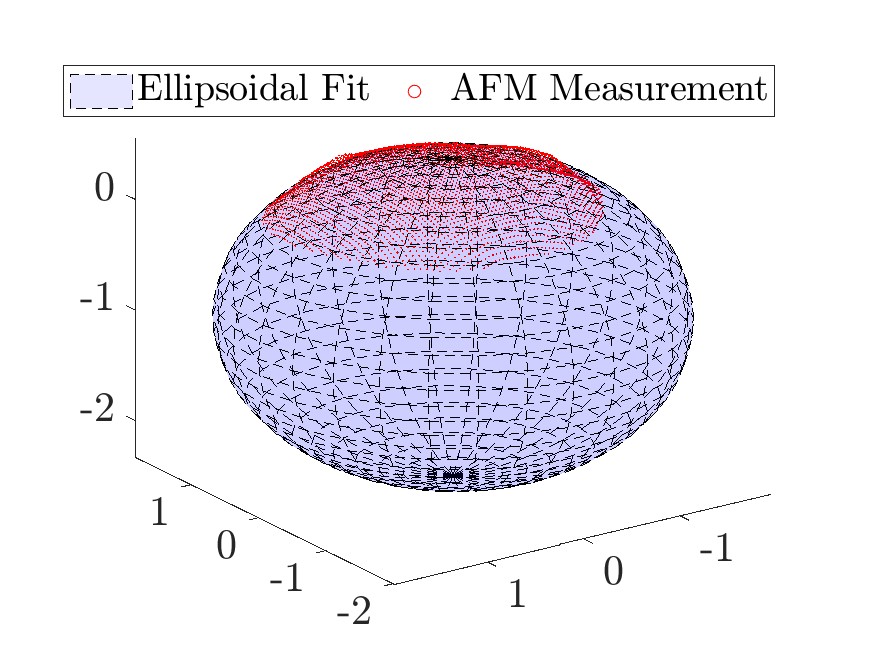} &
    \includegraphics[width=.3\textwidth]{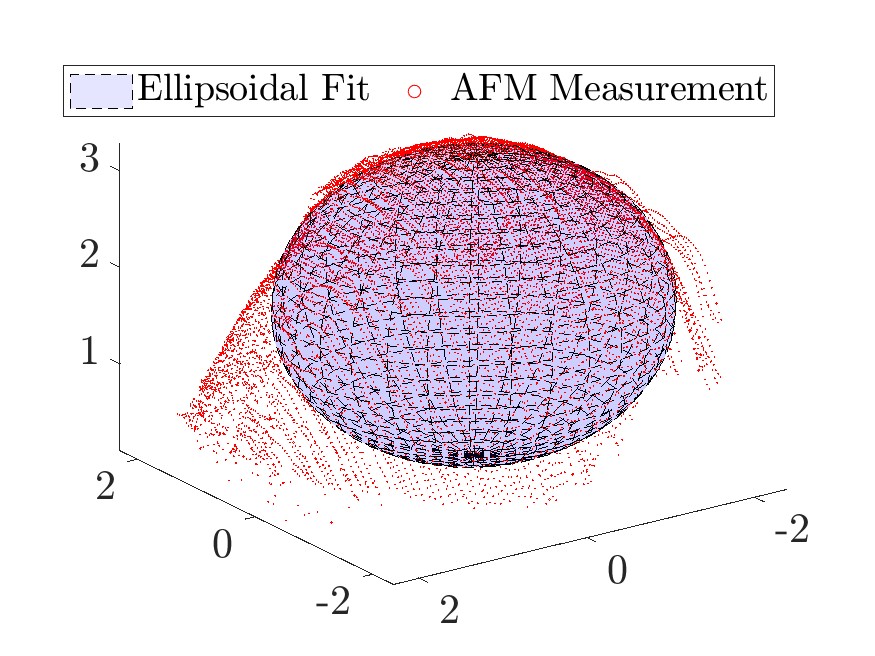} 
    \end{tabular}
    \caption{Visualization of ellipsoidal fits. Red points indicate the surface points measured by AFM. Top row: Guest particles before mechano-fusion. Bottom row: Guest particles after mechano-fusion.}
    \label{fig:fit_vis}
\end{figure}

This minimization problem is solved using the Matlab function \texttt{fmincon}, which employs an interior-point algorithm~\cite{byrd.2000}. 
Along the steep descent at the boundary of a particle, the interaction of the geometry of the cantilever and the geometry of the surface causes inaccurate measurements, called tip-sample convolution effects~\cite{Shen.2017}, see the right-hand column of Figure~\ref{fig:fit_vis}.
In order to minimize these effects, we disregard all measured points $w\in W$, whose $z$-value is below the $30$-th percentile of $z$-values in $W$. 
In this way we focus only on the more accurate measurements close to the peak of each particle that define their 3D shape.
The interior-point algorithm used for solving the minimization problem~\eqref{eq:min_problem_general} also requires an initial point for the parameter vector $(\cx,\cy,\cz,a,b,c,\gamma)$.
Recall that we already performed an initial centering on the points of $W$ so that we can use $\cx^{\text{init}}=0$ and $\cy^{\text{init}}=0$ as initial values for $\cx$ and $\cy$.
The initial value for $\cz$ is heuristically chosen as 
\begin{equation}
    \label{eq:c_z_init}
    \cz^{\text{init}}=\min\{z\in\R\colon (x,y,z)\in W\},
\end{equation}\emph{i.e.}, the smallest $z$-value observed in $W$. The initial value for $\gamma$ is $\gamma^{\text{init}}=0$.
For the remaining parameters, we use the solution to a simplified version of the problem as initial values. Here we consider a least squares approximation in the sense of surfaces, \emph{i.e.}, functions from $\R^2$ to $\R$. That is, we consider a parametric function $f_{r_1,r_2}\colon\R^2\longrightarrow\R$ of the form
\begin{equation}
    \label{eq:half-ellipsoid}
    f_{r_1,r_2}(x,y) = 
    \begin{cases}
        \frac{r_2\sqrt{r_1^2-x^2-y^2}}{r_1}+ \cz^{\text{init}},\qquad &\text{if }r_1^2-x^2-y^2\geq 0,\\
        \cz^{\text{init}},\qquad &\text{else},
    \end{cases}
\end{equation}
for some $r_1,r_2>0$. 
The first summand on the right-hand side of Eq.~\eqref{eq:half-ellipsoid} describes the $z$-values of a half-spheroid with radius $r_1$ in $x$- and $y$-direction and radius $r_2$ in $z$-direction.
The second summand shifts the half-spheroid in $z$-direction.
Thus, the coordinates of its center are the initial values $\cx^{\text{init}}$, $\cy^{\text{init}}$ and $\cz^{\text{init}}$.
Finally, a function of the form~\eqref{eq:half-ellipsoid} is fitted in a least squares sense to the surface measurement acquired by AFM, using the Matlab function \texttt{fit}, which yields values $\widehat{r_1}$ and $\widehat{r_2}$. We then put $a^{\text{init}}=\widehat{r_1}$, $b^{\text{init}}=\widehat{r_1}$ and $c^{\text{init}}=\widehat{r_2}$.

\section{Results and Discussion}
The methods described in Sections~\ref{sec:flattening} and~\ref{sec:segmentation} are applied to the images acquired by AFM to receive a particle-discrete segmentation of every measurement of PS guest particles, both before and after MF. The methods of Section~\ref{sec:ellipsoid_fit} are then applied to every region of the resulting segmentations. In this way we obtain two data sets of ellipsoidal fits corresponding to guest particle shapes before and after MF. In total,  $40$ guest particles before MF and $240$ guest particles after MF were considered.

\subsection{Quantification of guest particle deformation}
\label{sec:deformation}
In this section we use  the ellipsoidal fits, which have been obtained by solving the minimization problem stated in~\eqref{eq:min_problem_general},
in order to quantify the deformation effect on the guest particles during the mechano-fusion process.
To this end, we consider the so-called aspect ratio $\alpha\ge 1$ of an ellipsoid with semi-axes of lengths $a$, $b$ and $c$, which is defined as 
\begin{equation}
    \alpha =\frac{\max\{a,b,c\}}{\min\{a,b,c\}},
\end{equation}
being the ratio of the length of the major semi-axis to the length of the minor semi-axis.
The aspect ratio $\alpha$ is a scalar quantity that quantifies the deviation of an ellipsoidal shape from the spherical shape.
Note that an ellipsoid is a sphere if and only if it has an aspect ratio of $1$, and a higher aspect ratio corresponds to a larger deviation from a sphere, and hence a higher degree of deformation.
We compute the aspect ratio $\alpha$ for every ellipsoidal fit obtained by the minimization problem~\eqref{eq:min_problem_general} for both samples of guest particles, before and after mechano-fusion. 
This allows us to quantify the degree of deformation of the guest particles after mechano-fusion in comparison to pristine PS particles by considering a single scalar quantity. 

\begin{figure}[ht]
    \centering
    \begin{tabular}{cc}
     \includegraphics[width=.45\textwidth]{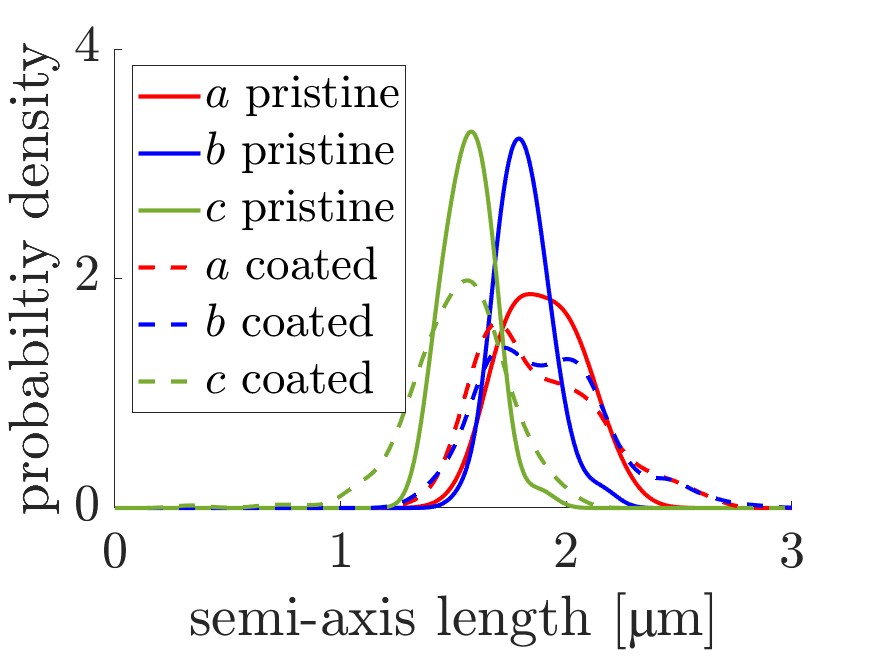} &
    \includegraphics[width=.45\textwidth]{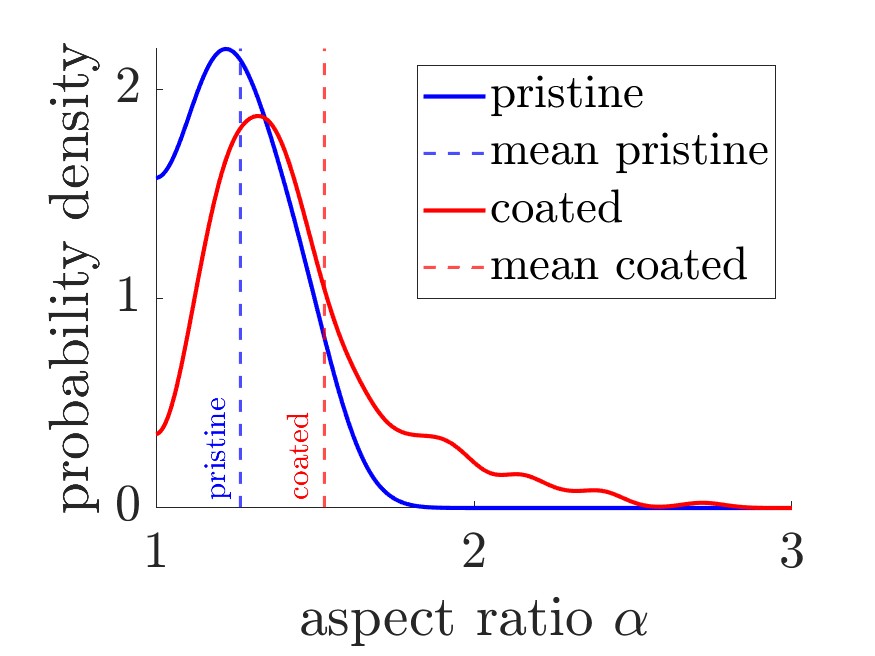} \\
    (a) & (b)
    \end{tabular}
    \caption{(a) Estimated probability densities of the three semi-axis lengths obtained through ellipsoidal fits before mechano-fusion (solid curves) and after mechano-fusion (dotted curves). 
    (b) Estimated probability densities of aspect ratios observed in samples of guest particles before mechano-fusion (blue) and after mechano-fusion (red). The dotted lines indicate the mean value of the corresponding distribution.}
    \label{fig:deformation}
\end{figure}

Figure~\ref{fig:deformation}a shows the probability densities, obtained by kernel density estimation,  of the three semi-axis lengths $a,b,c$ of the obtained ellipsoidal fits for guest particles before and after mechano-fusion. 
Note that it is not possible to directly compare the densities of $a$ and $b$ obtained for the pristine sample with those of the coated sample, since the rotation around the $z$-axis by the angle $\gamma$ can interchange the roles of $a$ and $b$ for each guest particle.
Nevertheless, we observe that MF leads to a clearly visible widening of the solid curves in Figure~\ref{fig:deformation}a , \textit{i.e.}, to
a strong increase of variance  of the distributions of all three semi-axis lengths. Additionally, the semi-axis lengths $a$ and $b$ tend to increase after MF, whereas a slight decrease in the semi-axis length $c$ is observed.
Figure~\ref{fig:deformation}b shows a comparison of the probability densities of  aspect ratio $\alpha$ before and after mechano-fusion. 
In particular, the solid curves show the  densities of $\alpha$, obtained by kernel density estimation. 
Here we can observe an overall shift towards higher aspect ratios.
Most notably, a portion of the guest particles after mechano-fusion exhibits strong deformations with  aspect ratios larger than $2$, whereas no aspect ratios larger than $2$ were observed in the pristine sample.
The dashed lines indicate the mean values of the corresponding distribution. 
In particular, the mean aspect ratio increases from $1.27$ before mechano-fusion to $1.53$ after mechano-fusion, which is an increase by $20.74 ~\%$.

\subsection{Assumptions on pristine guest particles}
\begin{figure}[ht]
    \centering
    \includegraphics[width=.6\textwidth]{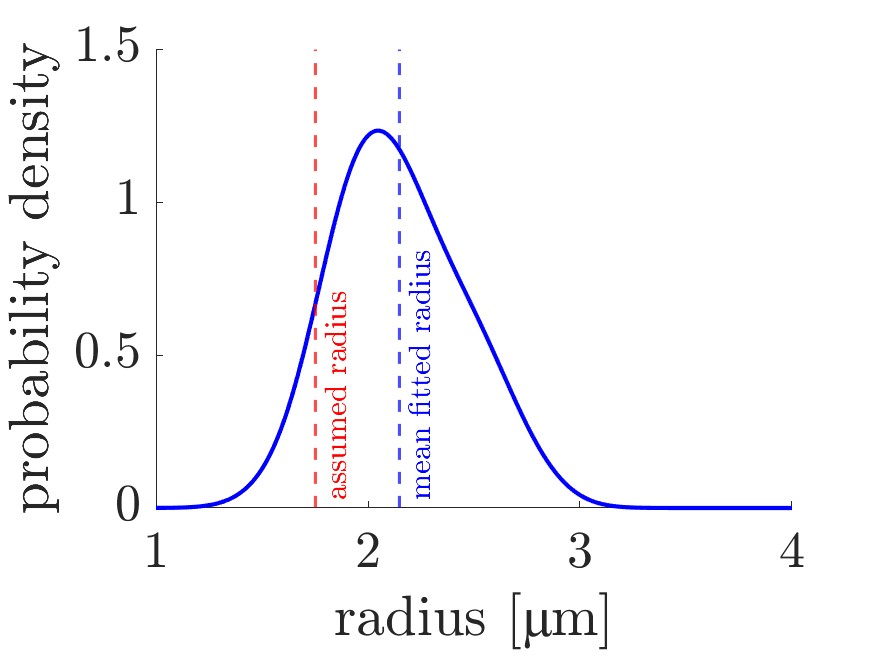}
    \caption{Estimated probability density of the radius $a$ of a spherical fit to the pristine particles without the volume constraint. The dashed blue line shows the mean value of the corresponding distribution, while the dashed red line shows the value assumed in Section~\ref{sec:ellipsoid_fit} for the radius of pristine guest particles.}
    \label{fig:radius_from_spherical_fit}
\end{figure}

The method used for reconstructing the shape of  guest particles by solving the minimization problem~\eqref{eq:min_problem_general} heavily relies on the assmuption that the volume of guest particles before mechano-fusion is known and that this volume does not change during mechano-fusion. 
In particular, the knowledge of the volume is obtained through the assumptions that pristine guest particles are spherical and monodispere with a radius of $\SI{1.75}{\micro\metre}$, see also Figure~\ref{fig:SEM-before}b.
However, the application of the  reconstruction  method  presented in this paper to measurements of guest particles before MF yields aspect ratios $\alpha$ that deviate notably from $1$, which is the aspect ratio of a sphere.
Hence, the assumption made regarding the volume of  guest particles does not lead to perfectly spherical fits for pristine guest particles. 
On the other hand, we can check if a spherical fit (instead of an ellipsoidal one) leads to the expected radius $R=1.75$ for pristine guest particles.
For this, we consider the unconstrained minimization problem 
\begin{equation}
    \label{eq:min_problem_spherical}
    \min_{\cx,\cy,\cz,a} H(\cx,\cy,\cz,a,a,a,0),
\end{equation}
where $H$ is given by Eq.~\eqref{eq:loss_function_def}. 
As we now require the lengths of all three semi-axis to be equal, the resulting shape described by Eq.~\eqref{eq:ellipsoid_def} is always that of a sphere. 
Note that we can set $\gamma=0$ since a sphere is invariant to rotations around its center. 
The minimization problem given in \eqref{eq:min_problem_spherical} is again solved by the Matlab function \texttt{fmincon}. 
The initial values of the vector $(\cx,\cy,\cz,a)$ for the optimization algorithm are chosen analogously to those described in Section~\ref{sec:ellipsoid_fit}, where only the value of $\widehat{r_1}$, which has been determined by means of Eq.~\eqref{eq:half-ellipsoid}, is used as an initial value for $a$.
Figure~\ref{fig:radius_from_spherical_fit} shows the probability density, obtained by kernel density estimation, of the radius $a$ after solving the minimization problem~\eqref{eq:min_problem_spherical} for all measured pristine guest particles. The dashed blue line shows the mean of fitted values for $a$, which is equal to $2.15$, whereas the dashed red line shows the expected radius $R=1.75$.

As we observed in Figure~\ref{fig:deformation}b, the assumption made on the volume of the PS guest particles did not lead to a perfectly spherical fit in the case of pristine guest particles before MF. Figure~\ref{fig:radius_from_spherical_fit} therefore shows an expected result, in the sense that the assumption of perfect sphericity does not lead to fits with a mean radius of $R=1.75$.
However, we would expect a rather small variance and deviation of the mean. 
The large variance of the estimated probability density shown in Figure~\ref{fig:radius_from_spherical_fit} and the significant deviation of the mean $2.15$ from the assumed value of $1.75$ do not agree with results from visual inspection of SEM images, see Figure~\ref{fig:SEM-before}b, which leads us to conclude that the assumption of perfect sphericity is not fully satisfied. In addition, concerning the results shown in Figure~\ref{fig:deformation}b, it is reasonable that even pristine guest particles before MF exhibit small impurities which can have a significant impact on the reconstruction of their 3D shape and hence their aspect ratio. 
This effect can also be augmented in the data by small inaccuracies in the measurements.

It is crucial to note that the violation of the sphericity assumption does not directly affect our method for reconstructing the 3D shape of the guest particles, as it only relies on the knowledge of their volume. 
Nevertheless, our method is able to quantify the guest particle deformation during MF through comparison of the aspect ratios before and after MF.

\section{Conclusion}
The method presented in this paper allows the reconstruction of the 3D shape of guest particles before and after MF by means of an ellipsoidal fit based on measurements obtained by AFM. 
The aspect ratio of these ellipsoids is used to quantify their deviation from a perfect spherical shape. 
By comparing the aspect ratios of the reconstructed shapes of guest particles before and after MF, we can thus quantify the degree of deformation experienced by guest particles during the MF process. 
The obtained results show a severe increase in the aspect ratios of guest particles after MF in comparison to pristine guest particles before MF, which corresponds to a larger deviation from a spherical shape, \emph{i.e.}, stronger deformation.
To the best of our knowledge, this is a novel use case of data acquired by AFM measurements. In particular, this opens the door for more involved investigations on the relationship between MF process parameters and the geometry of the resulting coated particles.
For example, the presented approach can be applied to different scenarios with varying process parameters or varying host and guest particles in order to study their individual influence on the resulting geometry of coated particles, and thus on their effective properties.

\section*{Acknowledgements}
We gratefully acknowledge funding by the German Research Foundation (DFG) under grants 
462365306 (SPP2289),
PE 1160/30-1
and SCHM 997/42-1. Furthermore, we thank Gert Schmidt  for acquiring SEM images used in this study.

\bibliographystyle{unsrt}
\bibliography{Literatur}

\end{document}